\documentclass[twocolumn, tighten, times, trackchanges]{aastex63}

\usepackage{graphicx}	
\usepackage{amsmath}	
\usepackage{amssymb}	
\usepackage{footnote}
\usepackage{tablefootnote}
\usepackage{booktabs}

\newcommand{\sedsub}{$\mathrm{_{SED}}$}
\newcommand{\oiisub}{$\mathrm{_{[O\,{\small II}]}}$}
\newcommand{\hbsub}{$\mathrm{_{H\beta}}$}

\newcommand{\sedsubbf}{$\mathbf{_{SED}}$}
\newcommand{\oiisubbf}{$\mathbf{_{[O\,{\small II}]}}$}
\newcommand{\hbsubbf}{$\mathbf{_{H\beta}}$}

\received{}
\revised{}
\accepted{}
\submitjournal{ApJL}

\shorttitle{Dynamical Effects of Stellar Feedback at $z\sim0.8$}
\shortauthors{Debora Pelliccia et al.}

\begin{document}

\title{Effects of Stellar Feedback on Stellar and Gas Kinematics of Star-Forming Galaxies at $0.6<z<1.0$}

\author[0000-0002-3007-0013]{Debora Pelliccia}\email{E-mail: dpelliccia@ucdavis.edu}
\affiliation{Department of Physics and Astronomy, University of California, Riverside, 900 University Avenue, Riverside, CA 92521, USA}\affiliation{Department of Physics, University of California, Davis, One Shields Avenue, Davis, CA 95616, USA}

\author{Bahram Mobasher}
\affiliation{Department of Physics and Astronomy, University of California, Riverside, 900 University Avenue, Riverside, CA 92521, USA}

\author{Behnam Darvish}
\affiliation{Cahill Center for Astrophysics, California Institute of Technology, 1216 East California Boulevard, Pasadena, CA 91125, USA}

\author[0000-0002-1428-7036]{ Brian C. Lemaux}
\affiliation{Department of Physics, University of California, Davis, One Shields Avenue, Davis, CA 95616, USA}

\author{ Lori M. Lubin}
\affiliation{Department of Physics, University of California, Davis, One Shields Avenue, Davis, CA 95616, USA}\affiliation{Visiting Scientist, The Observatories, The Carnegie Institution for Science, 813 Santa Barbara Street, Pasadena, CA 91101, USA}

\author[0000-0002-5368-8262]{Jessie Hirtenstein}
\affiliation{Department of Physics, University of California, Davis, One Shields Avenue, Davis, CA 95616, USA}

\author{Lu Shen}
\affiliation{CAS Key Laboratory for Research in Galaxies and Cosmology, Department of Astronomy, University of Science and Technology of China, Hefei 230026, China}

\author[0000-0002-9665-0440]{Po-Feng Wu}
\affiliation{National Astronomical Observatory of Japan, Osawa 2-21-1, Mitaka, Tokyo 181-8588, Japan}

\author[0000-0002-6871-1752]{Kareem El-Badry}
\affiliation{Department of Astronomy and Theoretical Astrophysics Center, University of California Berkeley, Berkeley, CA 94720}

\author[0000-0003-0603-8942]{Andrew Wetzel}
\affiliation{Department of Physics, University of California, Davis, One Shields Avenue, Davis, CA 95616, USA}

\author[0000-0001-5860-3419]{Tucker Jones}
\affiliation{Department of Physics, University of California, Davis, One Shields Avenue, Davis, CA 95616, USA}



\begin{abstract}
Recent zoom-in cosmological simulations have shown that stellar feedback can flatten the inner density profile of the dark matter halo in low-mass galaxies. A correlation between the stellar/gas velocity dispersion ($\sigma_{star}$, $\sigma_{gas}$) and the specific star formation rate (sSFR) is predicted as an observational test of the role of stellar feedback  in re-shaping the dark matter density profile. In this work we test the validity of this prediction by  studying a sample of star-forming galaxies at $0.6<z<1.0$ from the \mbox{LEGA-C} survey, which provides high signal-to-noise measurements of stellar and gas kinematics. We find that a weak but significant correlation between $\sigma_{star}$ (and $\sigma_{gas}$)  and sSFR indeed exists for galaxies in the lowest mass bin (M$_\ast\sim10^{10}\,$M$_\odot$). This correlation, albeit with a $\sim$35\% scatter, holds for different tracers of star formation, and becomes stronger with redshift. This result generally agrees with the picture that at higher redshifts star formation rate was generally higher, and galaxies at M$_\ast\lesssim10^{10}\,$M$_\odot$ have not yet settled into a disk. As a consequence, they have shallower gravitational potentials more easily perturbed by stellar feedback.
The observed correlation between $\sigma_{star}$ (and $\sigma_{gas}$)  and sSFR supports the scenario predicted by cosmological simulations, in which feedback-driven outflows cause fluctuations in the gravitation potential that flatten the density profiles of low-mass galaxies.

\end{abstract}



\section{Introduction}
Understanding galaxy formation in the context of the cosmological framework is still an open question in astrophysics. While the $\Lambda$ cold dark matter ($\Lambda$CDM) cosmological model successfully explains structure formation in the universe \citep[e.g.,][]{Spergel2007, Komatsu2011}, dark matter only simulations have raised problems for this model, especially on small scales. These N-body simulations predict steep (or `cuspy') dark matter inner density profiles \citep*[e.g.,][]{Navarro1997}; however, observations have shown that the dark matter profiles of low-mass galaxies can be shallower than the predictions \citep[e.g.,][]{Spano2008, Oh2011}.

Recent works showed that adding a baryonic component to cosmological simulations can resolve the disagreements between predictions and observations for low-mass galaxies \citep[\mbox{M$_\ast \leq 10^{9.5}\,$M$_\odot$}, see e.g.,][]{Navarro1996,ReadGilmore2005,Governato2012, PontzenGovernato2012, Chan2015}. Stellar feedback may be able to alter the dark matter distribution of dwarf galaxies through bursts of star formation and subsequent gas outflows, which displace enough mass to significantly flatten the central dark matter profile. Low-mass galaxies have shallow gravitational potentials, and therefore are especially sensitive to stellar feedback.

\citet{El-Badry2016, El-Badry2017}, using the Feedback In Realistic Environments \citep[FIRE;][]{Hopkins2014} simulations, showed that fluctuations in the gravitational potential following bursts of star formation cause strong fluctuations in stellar kinematics of low-mass galaxies (\mbox{M$_\ast \leq 10^{9.5}\,$M$_\odot$}). Moreover, they found that galaxy specific star formation rate (sSFR) and  line-of-sight stellar velocity dispersion ($\sigma_{stars}$) have similar time-evolution, e.g.,  $\sigma_{stars}$ is higher during episodes of higher sSFR, though with a delay of $\sim\,$50~Myr. 
This connected time-evolution and delay is interpreted as the galaxy gravitational potential (probed by $\sigma_{stars}$) becoming deepest when cold gas accumulates in the galactic center, which in turn drives high sSFR. Subsequently, when stellar feedback causes the cold gas in the center of the galaxy to be removed through galactic outflows, sSFR starts to decrease almost instantaneously, while $\sigma_{stars}$ decreases only when enough gas mass is displaced and the gravitational potential is significantly shallowed. A positive correlation between 100~Myr averaged sSFR and $\sigma_{stars}$ was observed in simulated galaxies by \citet{El-Badry2017}, who suggested that this correlation could be used as an observational test to determine whether real galaxies undergo feedback-driven potential fluctuations as shown in the simulations, and therefore, whether stellar feedback is able to regulate galaxy stellar and dark-matter densities.

Observational hints of this correlation were shown by \citet{Cicone2016} for a large sample of isolated galaxies with M$_\ast \lesssim 10^{10}\,$M$_\odot$ in the local universe, and by \citet{Hirtenstein2019} for a sample of gravitationally lensed low-mass galaxies (M$_\ast =10^{8}- 10^{9.8}\,$M$_\odot$) at $z\sim2$ using gas kinematics to trace the galaxy potential. Observational evidence of the effect of stellar feedback on stellar kinematics is still missing beyond the local universe.  At higher redshifts, galaxies have typically higher gas fractions at fixed stellar mass \citep[as much as five times, see e.g.,][]{Schinnerer2016}, and  experience stronger episodes of star formation \citep[e.g.,][]{MadauDickinson2014}, which could displace enough mass to alter the dark matter halo profile even for more massive galaxies. Moreover, at $z\gtrsim0.2$ galaxies with M$_\ast \lesssim 10^{10}\,$M$_\odot$ are thought not to have dynamically settled into a disk yet \citep[e.g.,][]{Kassin2012, Simons2015}, which could result into shallower potentials at fixed $M_{\ast}$, which are more readily perturbed by stellar feedback.
Zoom-in cosmological simulations by \citet{Maccio2012} and \citet{Mollitor2015} have shown that reasonably strong baryonic feedback can also alter the dark matter density profiles of the progenitors of Milky Way-mass galaxies (\mbox{M$_{\mathrm{tot}}\approx\,$10$^{12}\,$M$_\odot$}) at intermediate redshifts. Moreover,  \citet{Maccio2012} compared the dark matter density profiles of two Milky Way-like simulated galaxies, one with strong and one with weak stellar feedback, and found that only the one with stronger feedback was able to flatten the density profile. They showed that the flattening starts at intermediate redshifts ($z\approx1-2$), when strong star formation and the subsequent energy transfer from feedback in shallower gravitational potentials has the strongest effect.

To observationally explore this hypothesis we investigated whether stellar and gas kinematics and sSFR are correlated for a sample of star-forming galaxies at \mbox{$0.6<z<1.0$}. This sample was taken from the Large Early Galaxy Astrophysics Census \citep[LEGA-C;][]{vanderWel2016} survey, which provides high signal-to-noise stellar and gas kinematics measurements. This Letter is organized as follows. In Section~\ref{sec:data} we introduce the data and our sample selection. In Section~\ref{sec:ssfr} we describe the methods used to derive our sSFR measurements. The core of the analysis is presented in Section~\ref{sec:ssfr_sigma}. We discuss and summarize our results in Section~\ref{sec:conclusion}.

Throughout this Letter, we adopt a \citet{Chabrier2003} initial mass function and a $\Lambda$CDM cosmology with H$_0$ = 70 km s$^{-1}$, $\Omega_\Lambda$ = 0.7, and $\mathrm{\Omega_M}$ = 0.3. 

\begin{figure*}[t!]
\centering
\includegraphics[width=0.7\textwidth]{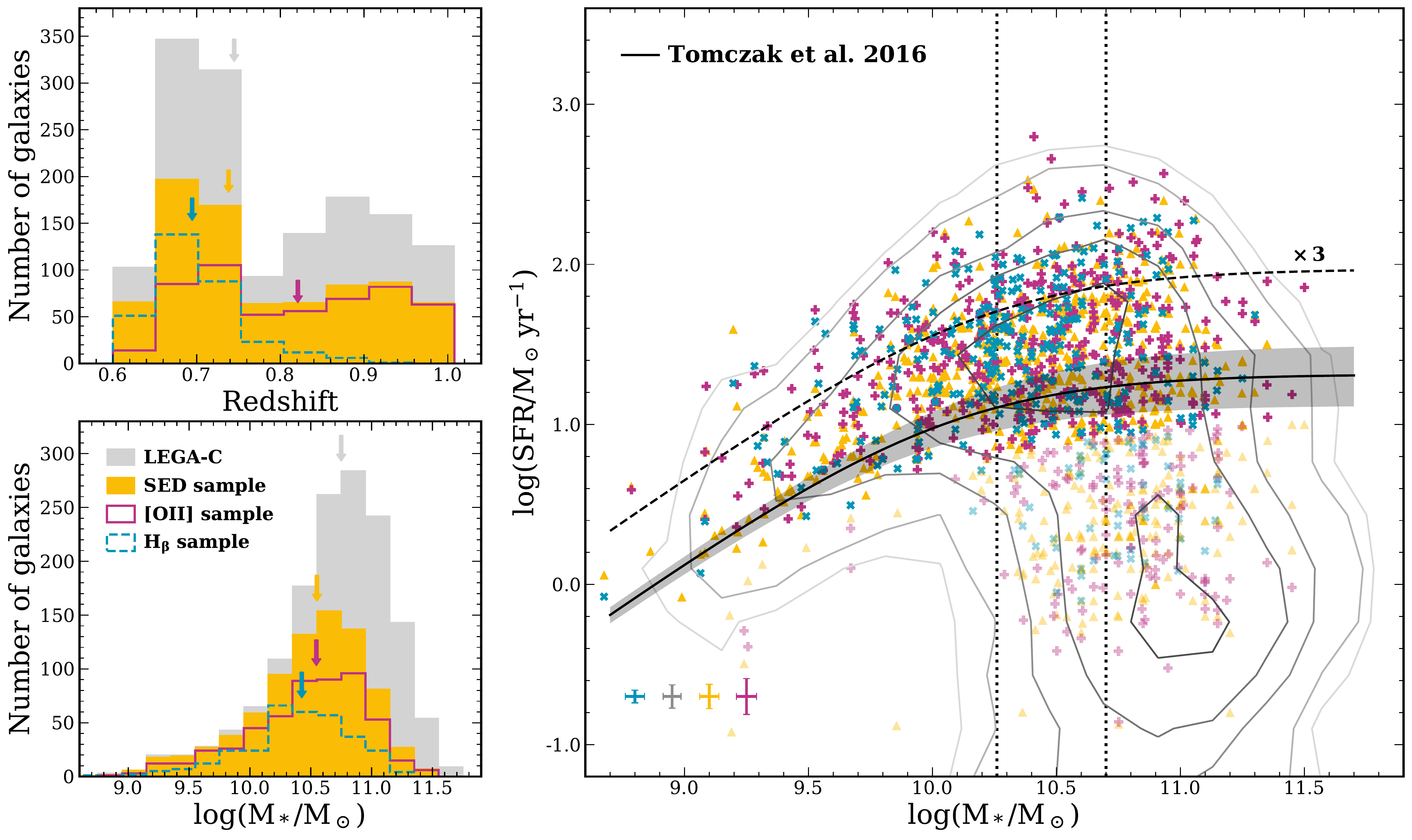}
\caption{Redshift (top-left panel) and stellar mass (bottom-left panel) distributions for the parent LEGA-C spectroscopic sample (gray) compared to the distributions for the star-forming SED (yellow), [\ion{O}{2}] (magenta), and H$_\beta$ (blue) samples (see Section~\ref{sec:data}). The arrows point to the median value of each distribution. \textit{Right-hand panel:} SFR vs M$_\ast$. Values of SFR\sedsub~are plotted for the parent LEGA-C (gray contours) and SED (yellow triangles) samples, while SFR\oiisub~and SFR\hbsub~(Section~\ref{sec:ssfr}) are plotted for  the [\ion{O}{2}] (magenta plus signs) and H$_\beta$ (blue crosses) samples, respectively. The median uncertainties in each sample are shown in the bottom-left corner. The contours show the 90\%, 70\%, 50\%, 30\% and 10\% of the density distribution of the parent  sample. The triangles, plus signs and crosses in lighter colors show the galaxies with SFR lower than $0.3\,dex$ below the relation (black line) for star-forming only galaxies at $z=0.8$ from \citet[][see discussion in Section~\ref{sec:data}]{Tomczak2016}. The shaded region shows the redshift evolution of SFR-M$_\ast$ relation between $z=0.6$ and $z=1.0$. The black dashed line shows the limit above which galaxies are classified as starbursts \citep[e.g.,][]{Elbaz2018}.  The black dotted vertical lines show the mass range of the M$_\ast$ bins used in the analysis in Section~\ref{sec:ssfr_sigma}. \label{fig:properties}}
\end{figure*}

\section{Data} \label{sec:data}
\subsection{LEGA-C Survey}  \label{subsec:legac}
The data used in this study are taken from data release~2 \citep[DR2;][]{Straatman2018} of the \mbox{LEGA-C} survey, which is a spectroscopic campaign carried out with the VIsible Multi-Object Spectrograph \citep[VIMOS;][]{LeFevre2003} on the ESO Very Large Telescope (VLT) aiming to study the stellar kinematics of $K_s$-band selected galaxies at \mbox{$0.6<z<1.0$} in the COSMOS fields. The $K_s$-band limit ranges from $K < 21.08$ at $z = 0.6$ to $K < 20.36$ at $z = 1.0$,  and results in a stellar mass limit of the order of $\sim$10$^{10}\,$M$_\odot$. We note that in the DR2 catalog, a small fraction ($\sim$20\%) of galaxies  are fainter than those limits, as they were observed to fill up the VIMOS masks \citep[see][for more details on the selection process]{vanderWel2016,Straatman2018}. We did not exclude those galaxies from our selection (described below); therefore, our sample has a small percentage of galaxies with stellar mass down to $\sim$10$^{9}\,$M$_\odot$. Observations of 20-hour integration have been performed using the high-resolution grism $HR-red$ (R = 2500) to obtain spectra over a wavelength range  \mbox{$\sim\,$6300$\,$\AA--$8800\,$\AA}, with spectral resolution (FWHM) of $\sim\,3\,$\AA~and typical signal-to-noise ratio (S/N)$\gtrsim$10~\AA$^{-1}$ on the continuum, required to extract the stellar kinematics. Examples of LEGA-C spectra are shown in \citet{vanderWel2016}  and \citet{Straatman2018}.

Measurements of the observed integrated gas and stellar velocity dispersions ($\sigma_{gas}$, $\sigma_{stars}$) are publicly released in DR2, and performed  by using the Penalized Pixel-Fitting \citep[pPXF,][]{Cappellari2004, Cappellari2017} code, which fit each  1D spectrum with a combination of high resolution stellar population templates and emission lines, downgraded to match the resolution of the LEGA-C spectra (see \citealt{Straatman2018} and \citealt{Bezanson2018} for a full description and examples of the pPXF fits). Values of $\sigma_{gas}$ and $\sigma_{stars}$ are measured as  the widths of the emission and absorption lines, respectively,  and represent the integrated velocity along the line of sight. This includes the contribution of both rotational and turbulent motions, tracing the underlying galaxy's total mass distribution (including stellar, gas, and dark matter components). These 1D kinematic measurements are not corrected for galaxy inclination and misalignment ($\Delta$PA) between the galaxy position angle (PA) and slit PA\footnote{As a LEGA-C observing strategy, the spectra were taken by placing the slits in the VIMOS masks always in a N-S direction \cite[see][]{vanderWel2016}.}, which could bias our results. We investigate the effect of inclination and $\Delta$PA on our results in Section~\ref{sec:ssfr_sigma}.

 In addition, when available,  LEGA-C DR2 provides measurements of fluxes (obtained by integrating the pPXF best-fit emission lines) and equivalent widths (EWs) for emission lines visible in the VIMOS spectral range ( i.e., [\ion{O}{2}]$\lambda$3727, H$\delta \lambda$4102, H$\gamma \lambda$4341,  H$\beta \lambda$4861,  [\ion{O}{3}]$\lambda$4959,  [\ion{O}{3}]$\lambda$5007), as well as spectroscopic redshifts ($z_{spec}$), Lick/IDS indices \citep{Worthey1997}, S/N measurements and quality flags \citep{Straatman2018}, which helped us to select a sample of galaxies with only high quality measurements (see Section~\ref{subsec:samp_select}).

\subsection{COSMOS2015 Catalog}  \label{subsec:cosmos2015}
Ancillary data are available from the COSMOS2015 photometric catalog \citep{Laigle2016}, which provides accurate spectral energy distribution (SED) fitting measurements, based on deep thirty-band UV--IR photometry that covers all galaxy types. From this catalog we used  measurements of stellar mass (M$_\ast$), star formation rate (SFR\sedsub{}), specific SFR (sSFR\sedsub{}), stellar color excess E(B-V)$_\ast$, and ``star-forming/quiescent'' classification. The latter, which is based on the NUVrJ color-color diagram \citep[see][for more detail]{Laigle2016}, was used to select only star-forming galaxies for our investigation of the possible correlation between sSFR and velocity dispersion (see Section~\ref{subsec:samp_select}). Although these measurements have been performed by fixing redshifts in the SED fitting to their photometric redshift ($z_{phot}$) values, we found that, in general, $z_{spec}$ and $z_{phot}$ agree well with a normalized median absolute deviation in \mbox{$|z_{spec}-z_{phot}|/(1 + z_{spec})$} of 0.008 and with only a few strong outliers: 0.4\% with \mbox{$|z_{spec}-z_{phot}|/(1 + z_{spec}) > 0.1$}. We are, therefore, confident that adopting the COSMOS2015 physical parameters at $z_{phot}$ is not degrading our analysis.

 \subsection{Sample Selection} \label{subsec:samp_select}
Out of the 1988 galaxies in the LEGA-C public catalog we discarded 368 galaxies
for which issues with the quality of the spectrum and/or data reduction were observed, redshift could not be measured, pPXF fit was clearly bad, or had issues with the interpretation of the measurements \citep[see][for more details on the quality flags]{Straatman2018}. Moreover, we restricted the sample to galaxies in the redshift range of $0.6<z_{spec}<1.0$ (galaxies at lower/higher redshifts were observed to fill the VIMOS masks), and removed 47 galaxies that presented spectral duplicates, reducing the sample to 1474 galaxies.

After cross-matching this LEGA-C selected sample with the COSMOS2015 catalog, we found that eight galaxies did not have a match and 69 galaxies are flagged as X-ray sources  \citep[see][for more detail]{Laigle2016} that suggested a possible AGN contamination. We decided to remove those galaxies, and after selecting only star-forming galaxies following the COSMOS2015 classification (see Section~\ref{subsec:cosmos2015}), the sample was reduced by $\sim40\%$, leaving us with 815 galaxies. 

As we will describe in the next section, we performed our investigation of possible correlation between sSFR and velocity dispersion by adopting three measures of star formation:  SED fitting, spectroscopic $H\beta$ and [\ion{O}{2}] fluxes. While our full sample (which we call ``SED sample'') of 815 galaxies have sSFR\sedsub{} measurements, for 30 of those galaxies $\sigma_{gas}$ is not available. Moreover, the full sample is further reduced when we  consider galaxies with available H$\beta$ and [\ion{O}{2}] fluxes at S/N$>$5, which is a cut that we apply to retain only high-quality flux measurements used for SFR estimates. Therefore, besides the ``SED sample'', we used in our analysis an ``[\ion{O}{2}] sample'' and ``H$\beta$ sample'' with 531 and 323 star-forming galaxies, respectively\footnote{An additional cut is included in the ``SED sample'' and ``[\ion{O}{2}] sample'', removing 12 and three galaxies, respectively, because they showed low-quality sSFR and/or $\sigma_{stars}$ measurements (i.e., relative errors larger than one). }.

The properties of our three sub-samples are shown in Figure~\ref{fig:properties}. In particular, the right-hand plot shows the SFR plotted against M$_\ast$, and we can see that at M$_\ast \lesssim10^{10}\,$M$_\odot$ all the three samples and the parent LEGA-C sample lack galaxies with low star formation, i.e., galaxies that are below the  SFR--M$_\ast$ relation typical for normal star-forming galaxies \citep[e.g.,][]{Tomczak2016}. This is a consequence of  LEGA-C being a magnitude-selected survey, and therefore, the lower mass galaxies that are bright enough to be observed are the ones with higher star formation. Because in our analysis (detailed in Section~\ref{sec:ssfr_sigma}) we investigated the possible correlation between sSFR and velocity dispersion in three different M$_\ast$ bins (whose boundaries are delineated by the dotted vertical line in Figure~\ref{fig:properties}), we decided to focus only on the \emph{upper} part of the SFR--M$_\ast$ relation in order to be able to make fair comparisons across all stellar masses.  We therefore removed from our sample all galaxies (shown by markers in lighter colors in the plot in Figure~\ref{fig:properties}) with SFR lower than $0.3\,dex$ below the SFR--M$_\ast$ relation for star-forming galaxies at $z=0.8$ from \citet{Tomczak2016}\footnote{Though the SFR-M$_\ast$ relation from  \citet{Tomczak2016} was constrained using UV+IR SFR estimates, we found that for our sample the differences between SFR$_{UV+IR}$, from \citet{Muzzin2013}, and SFR\sedsub~are minimal (median $\Delta$logSFR=-0.09 with scatter of 0.38\,$dex$). }. The value of $0.3\,dex$ was chosen to roughly take into account the redshift evolution of the relation and the average uncertainties on the SFR measurements. This further cut left us with the final ``SED sample'', ``[\ion{O}{2}] sample'' and ``H$\beta$ sample'' having 578, 399, and 267 galaxies, respectively (see Table~\ref{table:samples}). We note that these three samples have 120 galaxies in common.

\section{Specific Star Formation Rate Measurements} \label{sec:ssfr}
In order to investigate the correlation between sSFR and velocity dispersion, we adopted three measurements of sSFR: one that comes from SED fitting (from the COSMOS2015 catalog), which traces a galaxy's star formation integrated over a relatively long period of time ($\sim$100\,Myr), and two from  [\ion{O}{2}] and H$\beta$ derived SFR (SFR\oiisub{} and SFR\hbsub{}, respectively), which are more sensitive to recent star formation episodes ($\sim$10\,Myr timescale). We applied an internal extinction correction to the emission line fluxes from the LEGA-C catalog, based on the stellar continuum reddening calculated from the SED fitting, and the \citet{Calzetti2000} reddening curve. Following  \citet{Wuyts2013}, dust attenuation due to the gas ($A_{\mathrm{gas}}$) is derived from the SED fitting stellar continuum attenuation ($A_{\mathrm{SED}}$) using the relation: $A_{\mathrm{gas}} = 1.9\,A_{\mathrm{SED}} - 0.15 A_{\mathrm{SED}}^{2}$. This dust correction was derived by \citet{Wuyts2013} for a large sample of massive ($M_\ast>10^{10}\,$M$_\odot$) star-forming galaxies at $0.7<z<1.5$, and it is widely used in literature \citep[e.g.,][]{Stott2016, Pelliccia2017, Swinbank2017}. Moreover, it shows a good agreement with the dust attenuation measured from Balmer decrement for stellar masses similar to the ones probed in this study \citep{Price2014}. Typical values of attenuation that we measured for our samples are $A_{\mathrm{H\beta}} \sim 3.0\,$mag and $A\mathrm{_{[O\,{\small II}]}} \sim 3.6\,$mag.

\begin{table}[t]
\caption{Samples \label{table:samples}}    
\vspace{-18pt}
\scriptsize
\begin{flushleft}
\tabcolsep=0.27cm
\begin{tabular}{cccccc} 
\toprule    
& & $\mathbf{N_{beforecut}}$ & $\mathbf{N_{aftercut}}$ & \textbf{\% SFR cut} & $\mathbf{\tilde{z}}$\\ 
\midrule[0.1em]
\textbf{SED} & M$_{\mathrm{low}}$  & 229 & 215 &   6\%  & 0.73\\        
 & M$_{\mathrm{med}}$  & 281 & 206 & 27\%  & 0.76\\   
 & M$_{\mathrm{high}}$  & 293 & 157 &  46\%  & 0.87\\
\midrule[0.05em]
$\mathbf{H_\beta}$& M$_{\mathrm{low}}$  & 114  & 111 &  3\% & 0.69\\        
 & M$_{\mathrm{med}}$  & 130 & 106 & 18\% & 0.70 \\   
 & M$_{\mathrm{high}}$  & 79 & 50 &  37\%  & 0.70\\
\midrule[0.05em]
\textbf{\ion{O}{2}} & M$_{\mathrm{low}}$  & 155 & 149 & 4\% & 0.79\\        
& M$_{\mathrm{med}}$  & 177 & 137 & 23\% & 0.83\\   
& M$_{\mathrm{high}}$  & 196  &  113 &  42\% & 0.91\\ 
\bottomrule
\end{tabular}\\[3pt]
\end{flushleft}
\vspace{-6pt}
{\scriptsize $\mathrm{N_{beforecut}}$ and $\mathrm{N_{aftercut}}$ are the numbers of galaxies in each mass bin of each sample before and after the SFR cut discussed in Section~\ref{subsec:samp_select}. We also report the fraction of galaxies that have been removed by this cut, and the median redshift ($\tilde{z}$) in each sub-sample. }
\end{table}

We decided not to use higher order Balmer decrement (i.g., H$\gamma$/H$\beta$ or H$\delta$/H$\beta$) to estimate dust extinction, as fluxes for H$\gamma$ and H$\delta$ emission lines with S/N$>5$ are only observationally accessible (in conjunction with H$\beta$ fluxes) for $\sim$25\% of our sample of star-forming galaxies. Moreover, obtaining reliable flux measurements for those Balmer lines is challenging, given their intrinsic weakness and the combined effects of stellar absorption and dust extinction \citep[e.g.,][]{Moustakas2006}. For the small sample of galaxies for which we could measure the Balmer decrement, we found that $\sim$40\% showed H$\gamma$/H$\beta$ and H$\delta$/H$\beta$ ratios higher than the theoretical values \citep[ i.e., 0.47 and 0.26, respectively,][]{Osterbrock2006} and the majority of the remaining galaxies showed inconsistent extinction values measured from two Balmer ratios (in general lower extinction values from H$\delta$/H$\beta$). An overcorrection for stellar absorption of H$\gamma$ and H$\delta$ fluxes  could explain the observed behavior, considering the weakness of those lines (\mbox{EW(H$\gamma$)$\,\sim-3.3$\AA} and EW(H$\delta$)$\,\sim-1.7$\AA~on average) and the strength of the measured stellar absorption (Lick/IDS indices H$\gamma_{A}\sim\,$4\AA~and H$\delta_{A}\sim\,$5.6\AA~on average).  H$\beta$ flux could also be affected by the overcorrection of stellar absorption, but because it is a stronger line (median EW(H$\beta$)$\sim-6$\AA), we would expect the effect to be negligible (although the observed differences between SFR\oiisub{} and SFR\hbsub{} could be a result of that, see below).

SFR\oiisub{} and SFR\hbsub{} are measured following \citet{Kewley2004} and \citet[][assuming the theoretical H$\alpha$/H$\beta$  for Case B recombination]{Kennicutt1998}, respectively, after applying the conversion from \citet{Salpeter1955} to \citet{Chabrier2003} IMF. For SFR\oiisub{} we use Eq.~4 in \citet{Kewley2004}, which does not take into account the dependence on the metallicity, because we were only able to measure the oxygen abundance 12+log(O/H) using high-S/N ($>5$) emission line ratio  \citep[e.g., R$_{23}$, see ][]{Zaritsky1994}  for a small sample (15) of our galaxies.
We do not believe that these 15 galaxies can be representative of the entire sample used for our analysis; therefore, we avoided using an average value of 12+log(O/H) from this small sample to correct SFR\oiisub{}. We find a good agreement between SFR\oiisub{} and the SFR measured from SED fitting, with negligible bias (median $\Delta$logSFR=-0.001) and scatter of 0.3\,$dex$, which is a reflection of the different timescales probed by the two tracers of star formation, as well as uncertainties in both processes. SFR\hbsub{} is in less good agreement with SFR\sedsub{}, with median $\Delta$logSFR=-0.11 (scatter equal to 0.22\,$dex$), and SFR\oiisub{}, which is in general 0.16\,$dex$ lower.  As mentioned before, these differences may be due to an overcorrection of the underlying stellar absorption. However, at this point we are not able to draw a definite conclusion, and an investigation of these differences is beyond the scope of this work. For this reason, we decided to perform our analysis using the three tracers of star formation separately, allowing to compare the results. The value of sSFR\sedsub{} used in the analysis presented in the next section are taken from the COSMOS2015 catalog and are computed through SED fitting, while sSFR\oiisub{} and sSFR\hbsub{} are computed as SFR/M$_\ast$, using M$_\ast$ measurements from COSMOS2015.

\begin{figure*}
\centering
	\includegraphics[width=\textwidth]{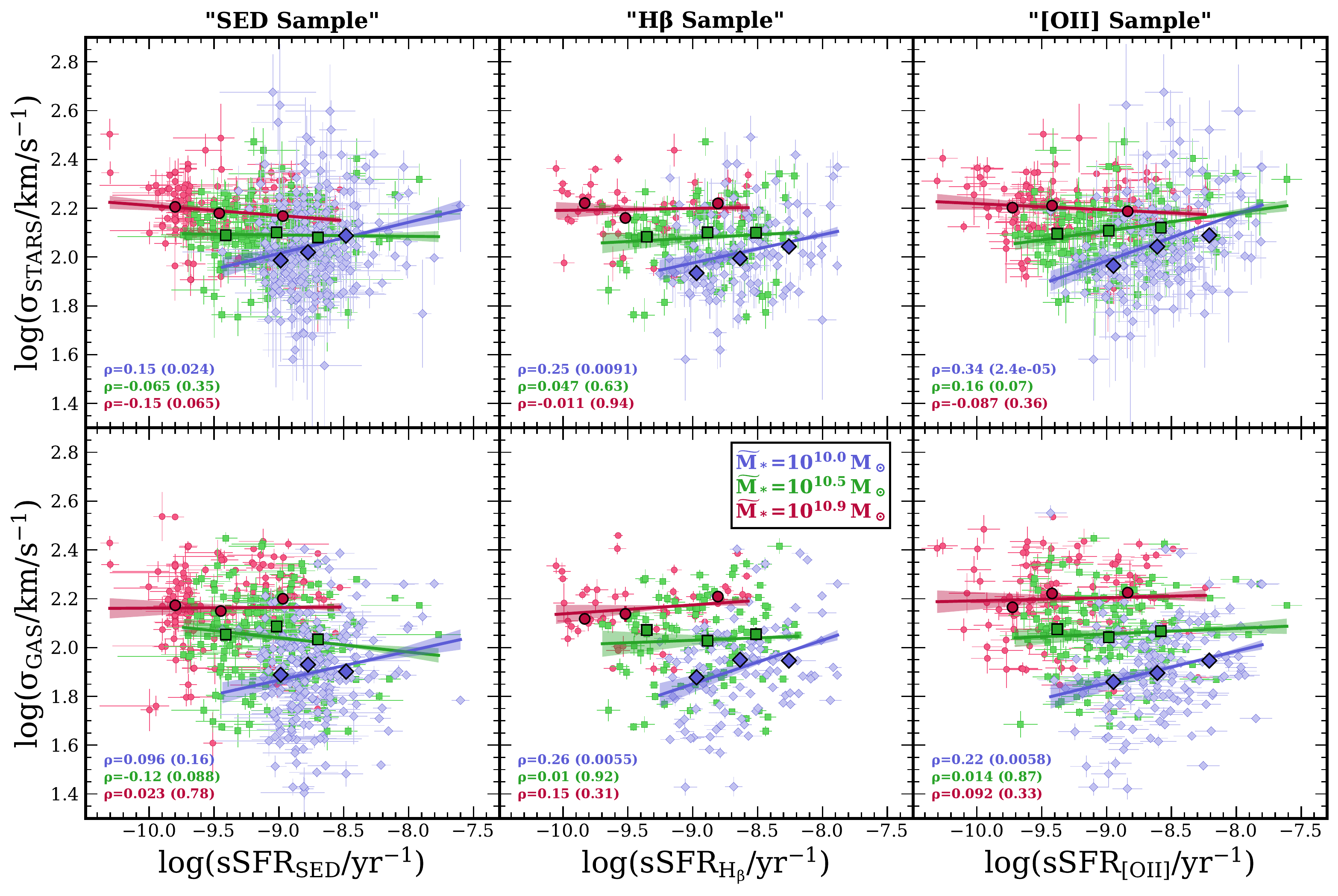}
    \caption{$\sigma_{stars}$ (top row) and $\sigma_{gas}$ (bottom row) as a function of the sSFR for the SED (left panels), H$\beta$ (middle panels), and  [\ion{O}{2}] (right panels) sample (see Section~\ref{sec:data}). Blue-violet diamonds,  green squares, and red circles represent galaxies with median M$_\ast=10^{10.0}$, $10^{10.5}$, and $10^{10.9}\,$M$_\odot$, respectively. The larger markers represent the median values of $\sigma_{stars}$ and $\sigma_{gas}$ in bins of sSFR with approximately equal number of galaxies per bin. The uncertainties on the median are smaller than the markers size ($\leq\,$0.03$\,dex$). The blue-violet, green, and red solid lines with shaded areas show the best-fit relations with 1$\sigma$ uncertainties for the galaxies in each mass bin (see Section~\ref{sec:ssfr_sigma}). The Spearman rank correlation coefficient in each mass bin is reported in the bottom-left corner of each panel, along with the corresponding two-sided $p$-value given in parenthesis (see Section~\ref{sec:ssfr_sigma}). }
    \label{fig:sigma_ssfr_all}
\end{figure*}

\section{sSFR - $\sigma$ correlation} \label{sec:ssfr_sigma}
In this section we investigate the possible correlation between sSFR and stellar/gas kinematics for a sample of star-forming galaxies at $0.6<z<1$.
Kinematics can be measured using different tracers, such as stars and neutral or ionized gas. Stellar and gas kinematics generally trace each other both in simulations and observations \citep[e.g.,][]{El-Badry2016, Bezanson2018}, with gas used as the preferred tracer because it is comparatively easier to detect due to the emission lines it produces.
However,  gas is \emph{directly} coupled to stellar feedback, and as a consequence gas
kinematics reacts to feedback on short timescales. Conversely, stars are not directly affected by feedback, but rather by the change in gravitational potential caused by feedback-driven outflows. Therefore, stellar kinematics is in principle a better tracer of the underlying potential, while gas kinematics can be affected by local turbulence caused by star formation. 
 The \mbox{LEGA-C} survey provides high quality measurements of both stellar and gas kinematics; therefore, we performed our analysis using both, allowing us to investigate if differences exist. This is the first time that an investigation of the relation between the galaxy integrated velocity dispersion and sSFR, predicted by the simulations, has been done using both stellar and gas kinematics at intermediate redshift.

\begin{table}[t]
\caption{Best-fit parameters of the $\sigma$--sSFR relations  \label{table:tab_fits}}    
\vspace{-17pt}
\scriptsize
\begin{flushleft}
\tabcolsep=0.13cm
\begin{tabular}{cccccc} 
\toprule    
& & \textbf{a} & \textbf{b} & \textbf{x$_0$}& \textbf{rms$_{\mathbf{intr}}$}\\ 
\midrule[0.1em]
$\mathbf{\sigma_{stars}}$& M$_{\mathrm{low}}$  & 2.042($\pm$0.012) & 0.128($\pm$0.044) & -8.78 & 0.144($\pm$0.011) \\        
\textbf{vs} & M$_{\mathrm{med}}$  & 2.091($\pm$0.009) & -0.006($\pm$0.025) & -9.02 & 0.110($\pm$0.008)  \\   
\textbf{sSFR\sedsubbf{}} & M$_{\mathrm{high}}$  & 2.190($\pm$0.009) & -0.041($\pm$0.022) & -9.47 & 0.097($\pm$0.008)   \\
\midrule[0.03em]
$\mathbf{\sigma_{gas}}$& M$_{\mathrm{low}}$  & 1.893($\pm$0.013) & 0.119($\pm$0.046) & -8.78 & 0.187($\pm$0.010) \\        
\textbf{vs} & M$_{\mathrm{med}}$  & 2.041($\pm$0.011) & -0.058($\pm$0.033) & -9.02 & 0.160($\pm$0.009)  \\   
\textbf{sSFR\sedsubbf{}} & M$_{\mathrm{high}}$  & 2.163($\pm$0.013) & 0.003($\pm$0.033) & -9.45 & 0.158($\pm$0.011)   \\
\midrule[0.1em]
$\mathbf{\sigma_{stars}}$& M$_{\mathrm{low}}$  & 2.018($\pm$0.015) & 0.116($\pm$0.046) & -8.64 & 0.136($\pm$0.014) \\        
\textbf{vs} & M$_{\mathrm{med}}$  & 2.081($\pm$0.013) & 0.028($\pm$0.036) & -8.88 & 0.125($\pm$0.011)  \\   
\textbf{sSFR\hbsubbf{}} & M$_{\mathrm{high}}$  & 2.195($\pm$0.016) & 0.008($\pm$0.035) & -9.52 & 0.103($\pm$0.014)   \\
\midrule[0.03em]
$\mathbf{\sigma_{gas}}$& M$_{\mathrm{low}}$  & 1.916($\pm$0.017) & 0.179($\pm$0.051) & -8.64 & 0.179($\pm$0.014) \\        
\textbf{vs} & M$_{\mathrm{med}}$  & 2.032($\pm$0.017) & 0.002($\pm$0.045) & -8.88 & 0.169($\pm$0.013)  \\   
\textbf{sSFR\hbsubbf{}} & M$_{\mathrm{high}}$  & 2.155($\pm$0.018) & 0.037($\pm$0.039) & -9.52 & 0.120($\pm$0.015)   \\
\midrule[0.1em]
$\mathbf{\sigma_{stars}}$& M$_{\mathrm{low}}$  & 2.056($\pm$0.013) & 0.191($\pm$0.036) & -8.63 & 0.120($\pm$0.014) \\        
\textbf{vs} & M$_{\mathrm{med}}$  & 2.109($\pm$0.010) & 0.074($\pm$0.024) & -8.98 & 0.100($\pm$0.010)  \\   
\textbf{sSFR\oiisubbf{}} & M$_{\mathrm{high}}$  & 2.204($\pm$0.011) & -0.025($\pm$0.024) & -9.41 & 0.099($\pm$0.010)   \\
\midrule[0.03em]
$\mathbf{\sigma_{gas}}$& M$_{\mathrm{low}}$  & 1.904($\pm$0.015) & 0.130($\pm$0.042) & -8.63 & 0.180($\pm$0.012) \\        
\textbf{vs} & M$_{\mathrm{med}}$  & 2.056($\pm$0.014) & 0.023($\pm$0.033) & -8.98 & 0.155($\pm$0.011)  \\   
\textbf{sSFR\oiisubbf{}} & M$_{\mathrm{high}}$  & 2.199($\pm$0.015) & 0.013($\pm$0.034) & -9.41 & 0.157($\pm$0.012)   \\
\bottomrule
\end{tabular}\\[3pt]
\end{flushleft}
\vspace{-6pt}
{\scriptsize The relations are expressed as $y = a + b(x - x_0)$, where $y$ represents log$\sigma_{star}$ or log$\sigma_{gas}$, $x$ is logsSFR and $x_{0}$ is the ``pivot'' value adopted to minimize the correlation between the errors on $a$ and $b$. M$_{\mathrm{low}}$, M$_{\mathrm{med}}$ and M$_{\mathrm{high}}$ refer to the three mass bins with median M$_\ast=10^{10.0}$, $10^{10.5}$, and $10^{10.9}\,$M$_\odot$. rms$_{intr}$ is the intrinsic scatter on the $y$ variable and is expressed in $dex$. }
\end{table}

We divided our SED, [\ion{O}{2}] and H$\beta$ samples (see Section~\ref{sec:data}) in three bins of stellar mass (M$_{\mathrm{low}}$, M$_{\mathrm{med}}$ and M$_{\mathrm{high}}$) with median values of $10^{10.0}$, $10^{10.5}$, and $10^{10.9}\,$M$_\odot$, and studied the correlation between the gas and stellar velocity dispersions ($\sigma_{gas}$, $\sigma_{stars}$) and sSFR in each M$_\ast$ bin. The mass range in each bin (delineated by vertical lines in the right-hand panel of Figure~\ref{fig:properties}) is chosen to be a compromise between having a significantly large number of galaxies and same median M$_\ast$ in the three samples (Table~\ref{table:samples}).
We proceeded with the analysis by fitting a relation between $\sigma$ and sSFR, adopting a linear least-squares approach \citep{Cappellari2013} that accounts for the uncertainties in both parameters and incorporates the measurement of the intrinsic scatter rms$_{intr}$ on the velocity dispersion (the best-fit parameters are reported in Table~\ref{table:tab_fits}). As an additional statistical tool  we use the non-parametric Spearman rank correlation test. This test provides two parameters, $\rho$ and the $p$-value. The parameter $\rho$, known as the correlation coefficient, provides information on the strength of the correlation. The $p$-value quantifies the significance of the correlation by giving the probability that the data are uncorrelated (i.e., the null hypothesis). We reject the null hypothesis for $p$-value$\,<\,$0.05.

The results of this analysis are presented in Figure~\ref{fig:sigma_ssfr_all}, where we also show the trends of the median $\sigma_{star}$ and $\sigma_{gas}$ in bins of sSFR with approximately equal number of galaxies per bin. 

We find that in general there exists a weak but significant correlation between $\sigma_{star}$ (and $\sigma_{gas}$) and sSFR for the galaxies in the M$_{\mathrm{low}}$ bin of all the three samples, except for \mbox{$\sigma_{gas}$ -- sSFR\sedsub{}}, where the correlation is not significant ($p$-value=0.16). This correlation appears to be strongest and most significant for \mbox{$\sigma_{star}$ -- sSFR\oiisub{}}, $\rho=0.34$ at $>$4$\sigma$ significance ($p$-value$\,=\,2.4e-05$), and somewhat less strong, $\rho\sim0.15-0.26$, but still significant ($\gtrsim\,$2.5$\sigma$) for the other samples. The best-fit relations show also a correlation at M$_{\mathrm{low}}$, and the slopes of such relations are consistent (within the uncertainties) with values found by \citep{Hirtenstein2019} using only $\sigma_{gas}$ for a sample of low-mass galaxies at $z\sim2$. We find that in general the intrinsic scatter rms$_{intr}$ around the relations in Figure~\ref{fig:sigma_ssfr_all} is fairly large ($\sim$35\% on average), with rms$_{intr}(\sigma_{gas})$  always larger than rms$_{intr}(\sigma_{star})$ (see Table~\ref{table:tab_fits}).
We do not find any significant correlations in the higher mass bins (M$_{\mathrm{med}}$ and M$_{\mathrm{high}}$) for all the samples, as shown by the Spearman rank correlation parameter $\rho$ and the $p$-value reported in the bottom-left corner of the plots in Figure~\ref{fig:sigma_ssfr_all}. The lack of significant correlation is also clear from the best-fit slopes reported in Table~\ref{table:tab_fits}, which have values close to zero for galaxies in M$_{\mathrm{med}}$ and M$_{\mathrm{high}}$, and  generally large uncertainties. 

The observed correlation is robust against jackknife re-sampling, the removal of galaxies with very high sSFR (sSFR$\,>10^{-8}\,$yr$^{-1}$), and the removal of galaxies with very low stellar mass \mbox{(M$_\ast<10^{9.4}\,$M$_\odot$)}.  Although we divided our samples in three mass bins to remove the dependence of sSFR and velocity dispersion on M$_\ast$, there may still be a residual dependence on M$_\ast$ within each mass bin, which could drive the observed relation. Therefore, we verified that within each mass bin, no trend with M$_\ast$ is observed  along the fitted relations, confirming that M$_\ast$ is not inducing the observed correlation. In addition, no clear trend with the attenuation is observed  along the fitted relation in  M$_{\mathrm{low}}$, excluding the attenuation as a possible driver of the observed correlation.
We also verified that the uncertainties on the estimates of M$_\ast$ did not bias our results as a consequence of using fixed bins of M$_\ast$. We created 1000 Monte-Carlo realizations of M$_\ast$ perturbed by its errors, and repeated the above analysis for each realization. After inspecting the mean and the spread of the posterior distributions of $\rho$, $p$-value and best-fit parameters, we found that the results presented here remain valid.

\citet{Bezanson2018} investigated the effect of galaxy inclination on $\sigma_{star}$ and $\sigma_{gas}$ for the LEGA-C galaxies. They found that, at fixed mass, $\sigma_{star}$ and $\sigma_{gas}$ can be underestimated by up to $\sim0.1-0.3\,dex$ for face-on star-forming galaxies, whereas this effect is not seen for all the other galaxy inclinations. We verified that face-on galaxies did not bias our results, by repeating our analysis after removing galaxies with inclination\footnote{ Both galaxy inclination (estimated using the galaxy axis ratio) and PA are taken from the COSMOS Zurich Structure and Morphology Catalog \citep{Scarlata2007}.}$\,<\,$40\degr. We found that our results are unchanged. Moreover, we investigated possible effects of PA misalignment ($\Delta$PA), by comparing values of $\sigma_{star}$ and $\sigma_{gas}$ to the average value in M$_{\mathrm{low}}$ as a function of $\Delta$PA. We found that only small ($\lesssim\,$0.06$\,dex$) and not statistically significant variations in $\sigma_{star}$ and $\sigma_{gas}$ are observed as a function of $\Delta$PA, which could not have meaningfully biased our results.

We are aware of the existence of a known large scale structure (LSS) at $z\approx0.72-0.76$ \citep{Scoville2007, Betti2019} in the COSMOS field. To verify that our results are not affected by this overdensity, we repeated the analysis removing the galaxies within the LSS redshift range. We found that our main results remain unchanged, and we still observe a significant correlation between $\sigma$ and sSFR for the galaxies in M$_{\mathrm{low}}$.

\subsection{Evolution with redshift}  \label{subsec:evolution}
As we can see from Figure~\ref{fig:properties} (left panel), the redshift distributions of the three samples show some differences (which are shown also in Table~\ref{table:samples}). While the SED and [\ion{O}{2}] samples are detected at every redshift between $z=0.6$ and $z=1.0$, the SED sample appears to be dominated by galaxies at $z<0.8$ and the [\ion{O}{2}] sample has a slightly stronger contribution from galaxies at $z>0.8$. Moreover, H$\beta$ emission detection drops rapidly at $z>0.8$, and as a result the vast majority of the galaxies in the H$\beta$ sample are at $z<0.8$. We investigated possible effects of the redshift evolution on the results presented here, by dividing the SED and [\ion{O}{2}] samples in two redshift bins ($z<0.8$ and $z>0.8$) and repeating the above analysis.  We excluded from this analysis the H$\beta$ sample, as it has only 19 galaxies at $z>0.8$.
We find that still no significant correlation exists for the galaxies in M$_{\mathrm{med}}$ and M$_{\mathrm{high}}$ at both redshifts. Moreover, the \mbox{$\sigma_{star}$ -- sSFR\oiisub{}} correlation for the galaxies in M$_{\mathrm{low}}$ is present both at $z<0.8$ and $z>0.8$, and the strength of this correlation increase with redshift, from $\rho=0.32$ ($\sim$3$\sigma$) at $z<0.8$ to $\rho=0.41$ (at $>$4$\sigma$) at $z>0.8$, though the increase in the slope of the fitted relation at $z>0.8$ is not statistically significant (see Table~\ref{table:tab_ros}). An evolution of the strength of the correlation is seen also for \mbox{$\sigma_{gas}$ -- sSFR\oiisub{}} and in the SED sample, where the correlations become too weak to be significant at $z<0.8$ (and the slopes of the fitted relations are close to $\sim$0), but become stronger and significant at $z>0.8$. Indeed, we found a significant ($\gtrsim$3$\sigma$) \mbox{$\sigma_{gas}$ -- sSFR\sedsub{}} correlation at $z>0.8$, which was not present for the overall sample. The values of the Spearman rank correlation parameter $\rho$ and the best-fit parameters for the galaxies in the lowest mass bin (M$_{\mathrm{low}}$) as a function of redshift are shown in Table~\ref{table:tab_ros}.

\begin{table}[t]
\caption{Redshift evolution of the correlation for galaxies in  M$_{\mathrm{low}}$  \label{table:tab_ros}}      
\vspace{-17pt}
\scriptsize
\begin{flushleft}
\tabcolsep=0.1cm
\begin{tabular}{cccccc} 
\toprule  
 & \textbf{a} & \textbf{b} & \textbf{x$_0$}& \textbf{rms$_{\mathbf{intr}}$} & $\mathbf{\rho}$ \textbf{(}$\mathbf{p}$\textbf{-value})\\ 
\midrule[0.1em]
\multicolumn{6}{c}{$\mathbf{\sigma_{stars}}$--\textbf{sSFR\sedsubbf{}} } \\ 
\midrule[0.03em] 
$\mathbf{z<0.8}$  & 2.031($\pm$0.014) & 0.082($\pm$0.050) &  -8.80 & 0.140($\pm$ 0.013) & 0.12 (0.17) \\        
\midrule[0.03em]  
$\mathbf{z>0.8}$  & 2.072($\pm$0.019) & 0.201($\pm$0.059) & -8.76 & 0.141($\pm$0.021) & 0.24 (0.018) \\   
\midrule[0.1em]
\multicolumn{6}{c}{$\mathbf{\sigma_{gas}}$--\textbf{sSFR\sedsubbf{}} } \\ 
\midrule[0.03em] 
$\mathbf{z<0.8}$  & 1.909($\pm$0.017) & -0.040($\pm$0.061) & -8.80 & 0.201($\pm$0.013) & -0.03 (0.74) \\        
\midrule[0.03em]  
$\mathbf{z>0.8}$  & 1.901($\pm$0.016) & 0.201($\pm$0.050) & -8.76 & 0.151($\pm$0.013) & 0.22 (0.029) \\   
\midrule[0.1em]
\multicolumn{6}{c}{$\mathbf{\sigma_{stars}}$--\textbf{sSFR\oiisubbf{}}} \\ 
\midrule[0.03em] 
$\mathbf{z<0.8}$  & 2.038($\pm$0.016) & 0.155($\pm$0.042) & -8.65  & 0.113($\pm$0.017) & 0.32 (0.0036)\\        
\midrule[0.03em]  
$\mathbf{z>0.8}$  & 2.086($\pm$0.017) & 0.234($\pm$0.045) & -8.61 & 0.119($\pm$0.020) & 0.41 (4.2e-05) \\   
\midrule[0.1em]
\multicolumn{6}{c}{$\mathbf{\sigma_{gas}}$--\textbf{sSFR\oiisubbf{}}} \\ 
\midrule[0.03em] 
$\mathbf{z<0.8}$  & 1.926($\pm$0.023) & 0.045($\pm$0.062) & -8.65 & 0.202($\pm$0.019) & 0.07 (0.52)\\        
\midrule[0.03em]  
$\mathbf{z>0.8}$  & 1.914($\pm$0.016) & 0.167($\pm$0.044) & -8.61 & 0.152($\pm$0.013) & 0.32 (0.0018) \\   
\bottomrule
\end{tabular}\\[3pt]
\end{flushleft}
\vspace{-6pt}
{\scriptsize Best-fit parameters (as described in Table~\ref{table:tab_fits}) and Spearman rank correlation parameter $\rho$, as well as the two-sided $p$-value, for the galaxies in the M$_{\mathrm{low}}$ bin as a function of redshift. See the discussion in Section~\ref{subsec:evolution}. }
\end{table}

\section{Discussion and Conclusions} \label{sec:conclusion}
In this Letter we tested the validity of a correlation between galaxy velocity dispersion and specific star formation rate, predicted by the simulation. Indeed, \citet{El-Badry2017} found that a correlation between $\sigma_{star}$ and sSFR exists for simulated galaxies with M$_\ast \lesssim10^{9.5}\,$M$_\odot$ that include prescriptions for stellar feedback. Given that these same simulations show that stellar feedback causes fluctuations in the gravitational potential, which in turn cause fluctuations in the stellar kinematics, the $\sigma_{star}$--sSFR relation can be used as an indirect probe of the effect of stellar feedback on the galaxy gravitational potential, and more generally of the role of stellar feedback in regulating the dark matter density profile of low-mass galaxies. Observational hints of this correlation were shown by \citet{Cicone2016} for a large sample of isolated galaxies with M$_\ast \lesssim10^{10}\,$M$_\odot$ in the local universe; however, at higher redshift the relationship between $\sigma_{star}$ and sSFR has not yet been investigated \citep[see][for a study of the \mbox{$\sigma_{gas}-\,$sSFR} relation at $z\sim2$]{Hirtenstein2019}.

We performed our analysis on a sample of star-forming galaxies at $0.6<z<1.0$, drawn from DR2 of the \mbox{LEGA-C} survey. Given the magnitude-selected nature of the survey, this sample is biased toward galaxies with high SFR at low stellar mass (M$_\ast\lesssim10^{10}\,$M$_\odot$). We decided, therefore, to focus our analysis only on galaxies that are above the \mbox{SFR-M$_\ast$} relation for normal star-forming galaxies, in order to make a fair comparison across all masses (see Section~\ref{subsec:samp_select}).

We have shown observational evidence of a weak but significant positive correlation between  $\sigma_{star}$ and sSFR for low-mass galaxies (M$_\ast\sim10^{10}\,$M$_\odot$) at $z\sim0.8$, in agreement with the predictions from cosmological simulations containing baryons. This correlation holds for different tracers of the star formation, i.e, for sSFR\sedsub{}, which traces star formation on $\sim$100\,Myr timescales, and for sSFR\hbsub{} or sSFR\oiisub{}, which trace star formation on $\sim$10\,Myr timescales. Theoretical prediction by \cite{El-Badry2017} showed a stronger correlation between $\sigma_{star}$ and sSFR averaged over the last 100\,Myr, as a consequence of $\sim$50\,Myr delay in the response of stellar kinematics to stellar feedback. Our analysis shows, instead, that the correlation is stronger for \mbox{$\sigma_{star}$ -- sSFR\oiisub{}}, e.g., for the shorter timescale star formation tracer, in agreement with the observational study at $z\sim2$ by \citet{Hirtenstein2019}. However, our observed correlations show a  relatively large scatter, which makes it difficult to draw a definitive conclusion about the observed differences in their strength.  A larger sample and a better characterization of the scatter is necessary to verify this result.
Our low-mass sample is slightly more massive than the galaxies predicted to feel the largest dynamical effects of stellar feedback in simulations (M$_\ast \lesssim10^{9.5}\,$M$_\odot$). However, a precise transition stellar mass is not specified at this point in simulations; therefore, we can confirm that the mass dependence of our result is \emph{qualitatively} consistent with the theoretical predictions.

We find that the positive $\sigma_{star}$--sSFR correlation becomes stronger with redshift (see Table~\ref{table:tab_ros}). This result is in agreement with \citet{Maccio2012}, who, using cosmological simulations, found that the effect of stellar feedback on galaxy gravitational potential is strongest at intermediate redshifts ($z\approx1-2$). Moreover, observations have shown that at those redshifts galaxies experience stronger episodes of star formation \citep{MadauDickinson2014} and that star-forming galaxies with M$_\ast\lesssim10^{10}\,$M$_\odot$ appear not to have settled in a disk yet \citep[e.g.,][]{Kassin2012, Simons2015}, possibly providing the right conditions for stellar feedback to effectively perturb the gravitational potential.

A significant correlation is also found between  $\sigma_{gas}$ and sSFR, although the scatter in this case is larger compared to $\sigma_{star}$ (see Table~\ref{table:tab_fits}). Similarly, \citet{Bezanson2018}, using LEGA-C data, found an increase of the scatter around the stellar mass Faber-Jackson relation \citep{FaberJackson1976} when using $\sigma_{gas}$ instead of $\sigma_{star}$, and attributed this effect to a large overall scatter existing between $\sigma_{gas}$ and $\sigma_{star}$ (0.13\,$dex$), while a good average agreement exists between the observed velocity dispersions. \citeauthor{Bezanson2018} stated that this agreement implies that $\sigma_{gas}$ can be used to constrain scaling relations for emission line galaxies, which is in line with what we found in this study; i.e., in absence of stellar kinematics measurements, gas kinematics can be used to test the effect of stellar feedback on the gravitational potential of galaxies. 
Finally, no significant correlation is observed between $\sigma_{star}$ (and $\sigma_{gas}$) and sSFR  for galaxies in the higher mass bins (M$_\ast\sim10^{10.5}\,$M$_\odot$ and M$_\ast\sim10^{10.9}\,$M$_\odot$). This result implies that more massive galaxies with deeper potentials are not affected by perturbations caused by stellar feedback, which is even weaker due to the decrease of gas fractions and the flattening of the SFR--M$_\ast$ relation at high M$_\ast$ \citep[e.g.,][]{Genzel2015}.

We did not attempt to directly compare the observed relation with the one predicted by the simulation, because at this point there are not publicly available measurements of sSFR and stellar kinematics for simulated galaxies with stellar masses comparable to the ones used in this Letter. This study was intended to explore whether a correlation between $\sigma_{star}$ (and $\sigma_{gas}$) and sSFR, which is predicted by simulations that incorporate realistic models of stellar feedback, indeed exists. By successfully showing that such a correlation does indeed exist for some sub-populations in our sample and not for others, we have provided observational results that can be used to constrain future theoretical predictions of stellar feedback.
A larger sample of galaxies is still necessary to better constrain this relation between stellar/gas velocity dispersion and specific star formation rate for low-mass galaxies, and characterize its dependence on other galaxy parameters. The final release of the LEGA-C survey will double the currently available measurements of  stellar and gas kinematics, allowing a deeper investigation of the dynamical effects of stellar feedback.   Comparisons between predicted and observed correlation could become a standard technique to constrain the feedback models adopted in simulations. Therefore, this relation, like others previously established (e.g., Tully-Fisher relation, mass-size relation) may constitute an essential benchmark for verifying theoretical models of galaxy formation and evolution.

\acknowledgments
Based on observations made with ESO Telescopes at the La Silla Paranal Observatory under program ID 194-A.2005 (The LEGA-C Public Spectroscopy Survey). D.P. acknowledges support from the NASA MUREP Institutional Opportunity (MIRO)  through the grant NNX15AP99A.
A.W. received support from NASA, through ATP grant 80NSSC18K1097 and HST grants GO-14734 and AR-15057 from STScI, the Heising-Simons Foundation, and a Hellman Fellowship.

\bibliographystyle{aasjournal}
\bibliography{legac_stellarfeedback2019}


\end{document}